\documentclass[10pt]{article}
\usepackage[utf8]{inputenc}
\usepackage[T1]{fontenc}
\usepackage{graphicx}
\usepackage{caption}
\usepackage[utf8]{inputenc}
\usepackage{booktabs}
\usepackage{subcaption}
\usepackage{amsmath}
\usepackage{setspace}
\usepackage[margin=1.0in]{geometry}
\usepackage{natbib}
\usepackage{longtable}
\usepackage{url}

\title{Learning and Evaluating Musical Features with Deep Autoencoders}

\author{Mason Bretan\\
\texttt{Georgia Tech}\\
\texttt{Atlanta, GA}\\
\and
Sageev Oore, Douglas Eck, Larry Heck \\
\texttt{Google Research}\\
\texttt{Mountain View, CA}\\
}

\date{June 2017}

\begin{document}

\maketitle

\begin{abstract}
In this work we describe and evaluate methods to learn musical embeddings. Each embedding is a vector that represents four contiguous beats of music and is derived from a symbolic representation. We consider autoencoding-based methods including denoising autoencoders, and context reconstruction, and evaluate the resulting embeddings on a forward prediction and a classification task.
\end{abstract}

\section{Introduction}
Music is a high dimensional and complex domain. Many of the tasks involving music analysis and information retrieval focus on reducing this dimensionality by categorizing music with discrete labels describing elements such as genre, composer, instrumentation, or era \cite{li2003comparative,jiang2002music}. However, new music is constantly being recorded, new genres are being identified/created, and so it is unreasonable to expect a musical dataset to have correct and comprehensive labelled examples for any of the above classifications. Any approach that depends on this will therefore not naturally scale.


In this article we focus on finding and leveraging the inherent continuity of the musical space. Instead of partitioning music based on a discrete labeling scheme we use deep neural networks to project music into a continuous, low dimensional space that captures meaningful characteristics. 
To assess the efficacy of this learned space, known as the ``embedding'', we first determine the properties that such an embedding should comprise.

Two common tasks associated with music are listening and composing. Thus, a useful musical embedding should have properties that allow it to successfully complete corresponding tasks: (1) ``make sense of what is heard'', i.e. extract meaningful features that correlate with human perceptions, and (2) ``compose'', i.e. complete a defined task related to music generation. We thus use two separate tests to measure the efficacy related to these tasks: composer identification, and generative prediction. Note that in fact, listening and composing are not at all independent of each other: real-world human musical activities such as performing, improvising, and playing together require significant integration of listening and composing, among other skills.

In this work we describe and evaluate a multitude of methods to learn the musical embeddings. In our context, each embedding is a vector that represents four contiguous beats of music (each of quarter note duration) and is derived from a symbolic representation. That is, our raw input data are MIDI piano rolls with accurate tempo information. The methods include networks trained to denoise, reconstruct context, predict forward, and discriminate between composers. In addition to the common training methods for completing these tasks we introduce an architecture and training methodology that influences the network to learn parameters that provide the embedding with a desired property.

\section{Related Work}
There is a considerable body of work that uses deep learning to extract semantic content from language-based data\cite{huang-et-al-2013,mikolov_efficient_2013,mikolov_distributed_2013,salakhutdinov-hinton-2007}. For example, Huang et al \cite{huang-et-al-2013} describe the use of a deep structured semantic model to new latent semantic models in order to project queries and documents into a common, lower-dimensional space in which cosine similarity provides a good estimate of the relevance between pairs of original items (i.e. before their projection). 

One clear intuition for finding such embeddings is that two textual excerpts can be very related even if they do not use the exact same keywords. The hope is that the mapping to a lower-dimensional, continuous, embedding does indeed take two such excerpts to nearby points. Similarly, two musical passages can be very related without using the exact same harmonic, melodic or rhythmic structure. We are thus interested in finding good embeddings that will achieve the same goal, within the context of music. Indeed, while music and language have significant structural parallels\cite{besson-2001}, it is generally only recently that researchers have begun to explore how this may give rise to cross-pollination between the respective fields.

In {\it ChordRipple}\cite{huang-et-al-2016}, Huang et al create a tool that helps composing students select a next chord in a sequence or replace a subsequence of chords within a longer progression. To achieve this they use an underlying semantic representation derived by learning embeddings of chords, so that chords are mapped to similar vectors when they occur near each other in the data. Madjiheurem et al \cite{madjiheurem-2017} compare the application of {\it skip-gram}-based models\cite{mikolov_distributed_2013} and {\it sequence-to-sequence} models\cite{sutskever_sequence_2014} for learning embeddings of chords. 

\section{Data and Representation}
\subsection{Dataset \& Augmentation}
In order to test the various networks and training procedures we used a collection of piano compositions from 25 artists. The distribution of compositions among artists is depicted in {\bf Table \ref{tab:dataset}}. At least two songs from each artist were held out for testing. Only songs with a time signature in which the note value of one beat is equal to a quarter note were used (i.e. the denominator of the time signature must have a value of 4). 

We augmented the data by transposing each piece into all keys. This also prevented the networks from simply learning the bias any composers might have had for specific key signatures.

\begin{table}[h!]
  \caption{Piano Music Dataset.}
  \label{tab:dataset}
  \begin{tabular}{ccl}
    \toprule
    Composer &  Num. Training Songs  & Num. Testing Songs\\
    \midrule
    Albeniz & 15 & 2\\
    J.S. Bach & 8 & 2\\
    Bartok & 21 & 4 \\
    Beethoven & 30 & 4\\
    Borodin & 8 & 2 \\
    Brahms & 31 & 5 \\
    Burgmueller & 10 & 2 \\
    Byrd & 34 & 4 \\
    Chopin & 49 & 6 \\
    Clementi & 17 & 2 \\
    Couperin & 10 & 2 \\
    Debussy & 10 & 2 \\
    Galuppi & 6 & 2 \\
    Grieg & 17 & 2 \\
    Handel  & 20 & 3\\
    Haydn & 20 & 3 \\
    Scott Joplin & 57 & 4\\
    Liszt & 17 & 2\\
    Mendelssohn & 16 & 2\\
    Mozart & 22 & 3\\
    Mussorgsky & 9 & 2\\
    Rachmaninov & 10 & 2\\
    Ravel & 5 & 2\\
    Scarlatti & 6 & 2\\
    Schubert & 30 & 4\\
    Schumann & 25 & 3\\
    Tschaikovsky  & 13 & 2\\
  \bottomrule
\end{tabular}
\end{table}

\subsection{Representation}
A piano roll representation of music can be thought of as an $m x n$ matrix where $m$ is the number of possible pitches and $n$ represents the total time. In this work we allow for 60 total pitches ranging from midi note 36 (two octaves below middle 'C') to 96 (three octaves above it). Any notes that fall outside of this range are transposed into either the lowest or highest octaves available.

Instead of using a standard decimal representation of time portraying seconds or milliseconds. Time is represented as ``ticks-per-beat'' (or ``pulses per quarter note'') which illustrate the smallest unit of time used for sequencing note events based on the musical notion of a beat. For example, if a value of four ticks-per-beat were used then the shortest possible note that could be accurately portrayed would be a sixteenth note. Triplets could not be accurately portrayed with this resolution. Typically 96 ticks-per-beat is considered adequate for capturing the temporal nuances of performance data and is the default resolution. The data used in this work come from quantized representations of a score (they are not performed) so capturing such nuances is not possible. Therefore, we use 24 ticks-per-beat which sufficiently captures the rhythms present in our data.

Each of the models described in the following section takes as input a $60 x 96$ matrix, corresponding to four beats (i.e. 96 ticks) worth of data. In the experiments only the note onsets are considered. The duration of each note is not addressed such that a matrix representing the beat sequence consisting of a quarter note middle `C' followed by three quarter note rests is identical to the matrix representing a beat sequence containing only a whole note middle `C'. Note onsets are represented with a `1' in the pitch-time location they occur and all other values in the matrix are `0'.

In order to train the following models a database of all segments of four contiguous beats in the training dataset are extracted. After the transposition augmentation process there are 4,682,293 four beat units available for training. We remove all duplicates and use 3,997,873 unique units for training the models described in the next section.

\section{Models}
In this section several methods for learning embeddings are presented. While the architecture and number of learnable parameters remain roughly the same for each model, the inputs and outputs are modified depending on the task. Each model involves a series of convolutional layers followed by fully connected layers leading to a 100-dimensional vector depicting the embedding. This size was chosen because it is small enough to quickly compute nearest neighbor search over a database consisting of over a million possible data points for real-time applications, yet large enough to embed many important musical features.

\subsection{Denoising Autoencoders}

A single layer in a neural network takes the output from the previous layer, ${\bf h}_{i-1}$, and applies an affine transformation followed by a nonlinearity to get ${\bf h}_{i} = f_{\theta_i} ( {\bf h}_{i-1} ) = \sigma_{i} ( {\bf W}_i {\bf h}_{i-1} + {\bf b}_i )$. We write $\theta_i = \{{\bf W}_i,{\bf b}_i \}$ to denote the parameters for layer $i$.

The first half of an autoencoder, the ``encoder'', consists of $n \geq 1$ such layers, $f_{\theta_1}, \ldots, f_{\theta_n}$, where the input to $f_{\theta_1}$ is simply ${\bf x}$, the input to the autoencoder. The output so far is given by
\[
{\bf h}_n = f_{\theta_n} ( f_{\theta_{n-1}} ( \ldots f_{\theta_1}({\bf x}) \ldots ))
\].

We also write ${\bf h}$ (i.e. with no subscript) to denote the output of the final (i.e. $n$th) layer of this encoder.

While the dimensionality of the intermediate layers may be larger than that of ${\bf x}$, 
the dimensionality of the final embedding, ${\bf h}$ is strictly (and usually considerably) less than that of ${\bf x}$. 
That is, if ${\bf x} \in R^d$ and ${\bf h} \in R^p$, then $p < d$.

The second half of the autoencoder, the ``decoder'' consists of the inverse mapping, 
\[
{\hat{\bf {x}}}_n = f_{\theta^{\prime}_{1}} ( f_{{\theta^{\prime}}_2} (\ldots f_{{\theta^{\prime}}_{n}}({\bf h}) \ldots ))
\]
where $f_{\theta'_i} = \sigma( {\bf W}'_i) + {\bf b}')$.


The network is trained to optimize the parameters $\theta = \{W,b\}$ by minimizing a given distance function. In order to capture local features that may repeat themselves, convolution is used so that the latent representation of the {\it k}-th feature map becomes
$h^k=\sigma(W^k + b^k)$.

To avoid simply learning the identity function the model learns to {\it denoise} a corrupted version of the input. In this work we test three methods of corruption:
\begin{enumerate}
    \item {\bf Random note drop} --- A random subset containing 50\% of the note onsets within a given input matrix are zeroed out and trained to be reconstructed.
    \item {\bf Random beat drop} --- All the note onsets within a randomly selected beat (a 60x24 space in the input matrix) are zeroed out and trained to be reconstructed.
    \item {\bf Octave split} --- Either the two lower octaves (24x96 space) or three upper octaves (36x96 space) are zeroed out and reconstructed. This is a rough estimate of splitting the piano roll into left and right hand parts, thus, the task for the network is to successfully reconstruct the right hand part given the left hand part and vice versa.
\end{enumerate}

To measure the distance between the reconstruction and original (non-corrupted) input, cosine similarity is used:

\begin{equation}
\it{sim}(\vec{X}, \vec{Y}) = \frac{\vec{X}^T\cdot \vec{Y}}{|\vec{X} || \vec{Y}|}
\end{equation}
where $\vec{X}$ and $\vec{Y}$ are two equal length vectors derived by flattening the reconstruction and ground truth matrices.

Negative examples are included using the following softmax function:
\begin{equation}
\it{\it{P}(\vec{R}|\vec{Q}) = \frac{\exp(\it{sim}(\vec{Q}, \vec{R}))}{\sum_{\vec{d} \epsilon D} \exp(\it{sim}(\vec{Q}, \vec{d}))}}
\end{equation}
where $\vec{R}$ is the flattened reconstructed matrix and $\vec{Q}$ is the non-corrupted input. $D$ is the set of five reconstructed matrices that includes $\vec{R}$ and four candidate reconstructed matrices derived from four randomly selected samples in the training set. The network then minimizes the following differentiable loss function using gradient descent:

\begin{equation}
\it{-log}\prod_{(Q, R)} P(\vec{R} | \vec{Q})
\end{equation}

\begin{figure*}[h!]
  \centering
  \includegraphics[width=.85\textwidth]{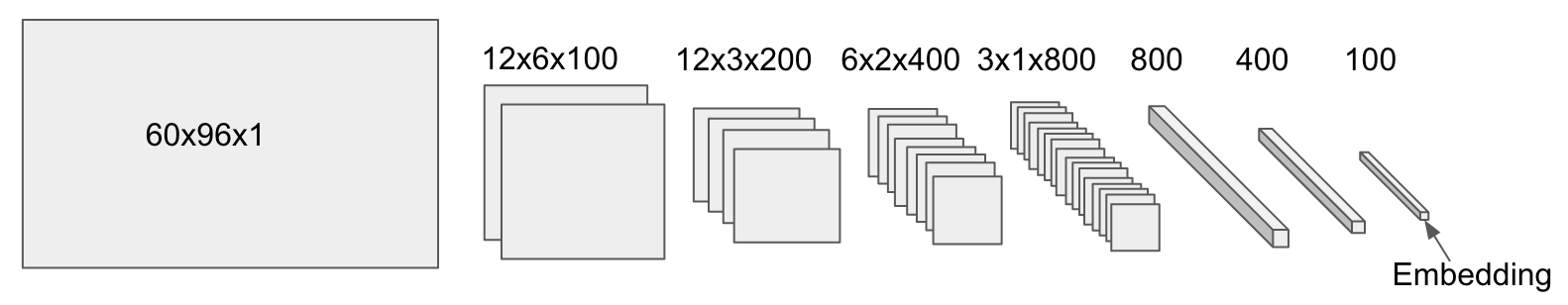}
  \caption{A denoising convolutional autoencoder is used to encode a piano roll representation of music. The encoder portion is depicted in the figure. The input is a $60 \times 96$ matrix consisting of four beats of music (24 ticks per beat) and 5 octaves worth of pitches. The first four hidden layers are convolutional and the following three are fully connected. The parameters of the encoder, $\theta = \{W,b\}$ and decoder, $\theta' = \{W',b'\}$ are constrained such that $W' = W^T$.
}
  \label{fig:music_cnn}
\end{figure*}

The architecture of the network is depicted in {\bf Figure \ref{fig:music_cnn}}. The first layer convolves a $12x6$ feature map (one octave of pitches and a sixteenth note duration worth of ticks) and has a stride rate of the same size. The subsequent convolutional layers use feature maps with stride rates that are half the size. Each layer uses exponential linear units (elu) and batch normalization is performed on each layer. The batch size is 100. The parameters of the encoder, $\theta = \{W,b\}$ and decoder, $\theta' = \{W',b'\}$ are constrained such that $W' = W^T$. A fully connected layer is used for the last layer of the encoder such that the embedding is a one-dimensional vector describing the entire measure.

\subsection{Context Prediction}
Predicting context has proven to be a powerful method for learning effective word embeddings. Similarly, we use the surrounding music of a specific four beat unit to learn the music embeddings. If $\bf U_i$ denotes the $i^{th}$ four beat unit in a sequence of four beat units from a single composition, the idea is to utilize the information provided by $\bf U_{i+1}$ and $\bf U_{i-1}$. Using an identical architecture as the previous autoencoder (including tied weights between encoder and decoder) we train two models with the tasks, respectively, of
\begin{enumerate}
    \item {\bf Forward prediction} --- The input to the network is $\bf U_i$ and the output is $\bf U_{i+1}$; and
    \item {\bf Contextual prediction} --- The input to the network is $\bf U_i$ and the output is $\bf U_{i-1} + U_{i+1}$. By summing the matrices of the surrounding units we maintain symmetry between the encoder and decoder, while obtaining 8 beats of context in a $60x96$ space.
\end{enumerate}

\subsection{Composer Classification}
Another task is to learn to classify four-beat units based on their composer. One rationale behind this task is that in order to be successful at predicting composer, the learned set of features has necessarily captured some important information about the musical passage. This is analogous to transfer learning methods used in computer vision tasks in which a network is first trained to label images and then, assuming relevant features have been learned, its parameters are used for additional tasks.

The architecture for the classification network is identical to the encoder portion of the previously described autoencoder. However, instead of a decoder, the 100-sized embedding vector is attached to a single output layer of size 27 (one for each composer). For the loss function we use softmax with cross-entropy.
 
\subsection{Regularized Training}
\begin{figure}[h!]
  \centering
  \includegraphics[width=.3\textwidth]{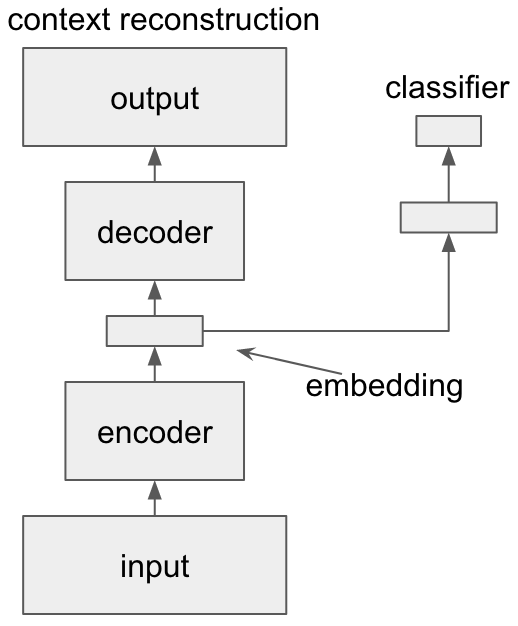}
  \caption{The elements of the network consist of the encoder, embedding, and decoder. The primary objective of the network is to reconstruct the surrounding context of a single four beat unit. An auxiliary task of identifying the composer of the input unit is included.}
  \label{fig:aux_training}
\end{figure}

In the last model we use the task of composer classification as a means to regularizing the embedding of the previously described contextual prediction network. The general premise is depicted in Figure~\ref{fig:aux_training}. Given the original encoder/decoder network an additional network is attached to the embedding layer. In particular, given a unit ${\bf U}_i$, we use its 100-dimensional embedding representation ${\bf h}({\bf U})$  as the (fully connected) input to a hidden layer with 50 hidden units, which in turn is fully connected to a 27-unit softmax layer.

The main objective of the network remains the same, given $\bf U_i$ then reconstruct $\bf U_{i-1} + U_{i+1}$. The auxiliary task of composer classification is imposed on the embedding such that network's parameters are optimized to not only reconstruct context, but also predict composer.

\section{Experiments and Results}

We use several techniques and measures to assess the efficacy of our embeddings.

\subsubsection{Forward Prediction \& Ranking} 
We train 3 stacked LSTM cells, each with 400 units, to predict the next step of a sequence. Specifically, at each time step, the input to the network is a the 100-dimensional embedding vector of a 4-beat unit ${\bf U}_i$. We train this network to predict the $8^{th}$ unit, ${\bf U}_{i+7}$ given the previous 7 units ${\bf U}_i, {\bf U}_{i+1}, \ldots {\bf U}_{i+6}$. This means that 28 beats of context (in four beat chunks) are provided before the prediction is made.

For each given target ${\bf U}_{i+7}$ as described above, we create a set of 1000 embedding vectors: one is ${\bf h}({\bf U}_{i+7})$, the true embedding for the target. The other 999 vectors are embeddings of randomly selected units in the data set.

Cosine similarity is measured between the output of the LSTM and the encodings of each unit. The similarities are then sorted and ranked from highest to lowest similarity (so the lower the rank the better).

The LSTM was trained with sequences provided by compositions in the training set and tested on 325,219 sequences from compositions in the test set. Embeddings from each of the previously described models were used resulting in seven prediction networks.

\subsubsection{Composer Classification}
In this experiment we test how useful the features learned in the 100-sized embedding vector are for discriminating between composers. To do this we train a neural network that takes in the embedding as an input, has a single elu activated hidden layer with 50 units, and an output layer of size 27 representing each artist. Given an input ${\bf U}$, the softmax classification layer outputs a probability distribution $P( C_j | {\bf U})$ over the set of composers $C_1, \ldots, C_{27}$. The composer with the highest probability, $f(x) = \arg\max_{i \in 1 \ldots27 } p_{i}(x)$, is used to predict the composer. 

\subsection{Results}
\begin{figure}[h!]
  \centering
  \includegraphics[width=.45\textwidth]{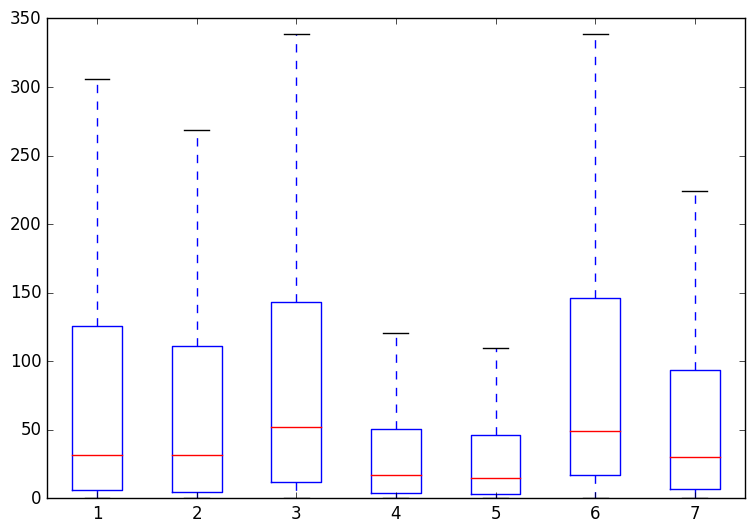}
  \caption{Boxplot of ranking results for each of the models: 1) Denoising autoencoder using random note drop 2) Denoising autoencoder using random beat drop 3) Denoing autoencoder using octave split 4) Next unit prediction 5) Context reconstruction 6) Composer discrimination and 7) Context reconstruction with composer regularization.
}
  \label{fig:rankings}
\end{figure}

For the prediction task we collected the ranks provided by each of the seven prediction LSTM networks for all 325,219 runs. We used these collection of ranks to compare models. The median, spread, and skew are reported in Table~\ref{tab:prediction} and Figure~\ref{fig:rankings} shows a boxplot of the results. Given that ranking distributions skewed considerably toward zero (the highest possible rank) we used the median instead of the mean.

To test for significance between the ranking distributions of each model t-tests comparing each possible model combination (21 possible comparisons) were performed. After a Bonferroni correction statistical significance is indicated by $p<0.00037$. All distributions are significantly different from each other.

To compare the spreads of each distribution we used Levene's test, which tests the null hypothesis that all input samples are from populations with equal variances. For all seven distributions the test statistic [W=7398.17] indicated significant differences at the $p<.001$ level.

\begin{table}[h!]
  \caption{Prediction}
  \label{tab:prediction}
  \begin{tabular}{cccc}
    \toprule
    Model &  Median Rank@1000  & Spread & Skew\\
    \midrule
    Autoencoder - Note Drop & 32 & 120 & 2.18\\
    Autoencoder - Beat Drop & 32 & 106 & 2.51\\
    Autoencoder - Octave Split & 52 & 131 & 2.22\\
    Predict Forward & 17 & 47 & 4.0 \\
    Predict Context & 15 & 43 & 3.98\\
    Composer Discrimination & 50 & 129 & 1.92\\
    Regularized & 30 & 87 & 2.31\\
  \bottomrule
\end{tabular}
\end{table}

The performance of the embeddings in the classification task are reported in Table~\ref{tab:classification}. For multi-class classification we computed two measures. The Micro-F1 score calculates metrics globally by counting the total true positives, false negatives and false positives. The Macro-F1 score calculates metrics for each composer separately and takes their average, weighted by the number of true instances for each composer. Thus, the Macro-F1 score accounts for label imbalance.
\begin{table}[h!]
  \caption{Composer Classification}
  \label{tab:classification}
  \begin{tabular}{ccc}
    \toprule
    Model &  Micro-F1  & Macro-F1\\
    \midrule
    Autoencoder - Note Drop & .14 & .48 \\
    Autoencoder - Beat Drop & .13 & .47\\
    Autoencoder - Octave Split & .12 & .43 \\
    Predict Forward & .14 & .52\\
    Predict Context & .14 & .53\\
    Composer Discrimination & .29 &.76 \\
    Regularized & .21 &.66\\
  \bottomrule
\end{tabular}
\end{table}

\section{Conclusions}

We have demonstrated and compared a variety of deep autoencoder architectures and objective functions to learn embeddings of note-based musical data. The network trained to reconstruct context performed the best at the LSTM prediction task. Unsurprisingly, the network trained to identify composers performed much better at this task than the features learned by the resulting embeddings of the denoising autoencoders and context prediction networks. However, by regularizing the contextual prediction network with this auxiliary classification task performance was improved. This network learned parameters that found a performance balance between the two experimental tasks.

\selectfont
\bibliography{music_embeddings}
\bibliographystyle{plain}

\end{document}